%% file: spinodal_prf.tex
\documentclass{jfm}
\usepackage{natbib,epsfig,graphicx,amsmath,amssymb,tikz}
\usepackage{mathrsfs}
\usepackage{bm}
\usepackage{color}

\input{symdef.inc}

\shorttitle{Kinetic energy spectra in binary-fluid mixtures}
\shortauthor{P. Perlekar}

\begin{document}
\title{Kinetic energy spectra and flux in turbulent phase-separating symmetric binary-fluid mixtures}
\author{Prasad Perlekar \aff{1} 
\corresp{\email{perlekar@tifrh.res.in}}}
\affiliation{\aff{1} Tata Institute of Fundamental Research,  Centre for Interdisciplinary Sciences, Hyderabad-500107, India.}
\maketitle
\begin{abstract}
We conduct direct numerical simulation (DNS) of the Cahn-Hilliard-Navier-Stokes (CHNS) equations to investigate the statistical properties of the turbulent phase-separating symmetric binary-fluid mixture.  Turbulence causes an arrest of the phase-separation which leads to the formation of a statistically steady emulsion. We characterize turbulent velocity fluctuations in an emulsion for different values of the Reynolds number and the Weber number.  Our scale-by-scale kinetic energy budget analysis shows that the interfacial terms in the CHNS provide an alternate route for the kinetic energy transfer.  By studying the probability distribution function (PDF) of the energy dissipation rate, the vorticity magnitude, and the joint-PDF of the velocity-gradient invariants we show that the statistics of the turbulent fluctuations do not change with the Weber number.
\end{abstract}

\section{Introduction}
\label{intro}

Below the consolute temperature, a symmetric binary-mixture in spatially homogeneous composition spontaneously phase separates forming domains of individual phases via the process of spinodal decomposition, see e.g. the books  \cite{chai98,gol05}. During the dynamics, individual domains merge and coarsen to form even larger domain until a final configuration is reached wherein two single-component domains separated by an interface is formed. The exact mechanism of domain growth depends on the interplay of viscous, inertial, and surface tension forces \citep{lif61,sig79,fur85,bra94,ken00,ken01,Puri09,datt15,cat18}.  

Presence of an external stirring such as shear or turbulent mixing counteracts the phase separation by breaking the coarsening domains to form a statistically stationary emulsion.  In presence of an external shear (with rate $\dot{\gamma}$),  the  size of a typical domain $D$ in an emulsion state can be estimated by the balance of shear stress $\eta \dot{\gamma}$  with the capillary force density $\sigma/D$ which gives $D\sim \sigma/{\eta \dot{\gamma}}$ (\cite{has95}). Here $\eta$ is the dynamic viscosity of the emulsion and $\sigma$ its surface tension. Although both experiments  \citep{Onu02} and numerical simulations \citep{sta06,str07,cat18} confirm formation of an emulsion state, the domain size shows different scaling with shear rate in the direction perpendicular and parallel to the shear \citep{sta06,str07,cat18}.  

The situation is much more tractable when the binary-fluid mixture emulsion is formed by an external stirring that generates homogeneous, isotropic turbulence (HIT). The average domain size $D$ of such  an emulsion can be estimated by a balance of inertial stress [$\rho  (\delta_D U)^2$]  with  the capillary force density ($\sigma/D$) \citep{hin55}, where $\delta_D U$ is the typical velocity difference across the domain. Using $\delta_D U \sim D^{1/3} \epsilon_\nu^{1/3}$ \citep{kol41,Fri96}, where $\epsilon$ is the energy dissipation rate,  provides an estimate for the average domain size as $D \sim  (\rho/\sigma)^{-3/5} \epsilon^{-2/5}$.  Early experiments \citep{pine84,eas92} indicated an arrest of the phase separation process in presence of HIT. This was  theoretically understood by using eddy diffusivity arguments by \citet{aro84}. However, only recent numerical investigations using a multicomponent Lattice-Boltzmann method in three-dimensions \citep{per14} and direct numerical simulations (DNSs) of Cahn-Hilliard-Navier-Stokes equations in two-dimensions \citep{ber05,fan16,per17,fan18} have been able to study emulsification by turbulence in symmetric binary-fluid mixtures.  Surprisingly, unlike shear flows, here numerically calculated domain size is in excellent agreement with  Hinze's prediction in  both  two- and three-dimensions.

It is apriori unclear whether turbulence in binary-fluid mixtures is similar or different than a single-component Newtonian fluid as only a small fraction of the volume is occupied by the emulsion interfaces. In two-dimensions, \citet{per17} show that the inverse energy cascade and the corresponding energy flux gets blocked at a wave-number corresponding to the domain size $D$.  In three-dimensions, numerical investigations   \citep{ken01,per14} show that for a binary-fluid mixture energy content in the inertial range is suppressed in comparison to a single-component fluid at the same Reynolds number. However, an understanding of the underlying energy transfer mechanisms in three-dimensions remains unclear. 

In this paper, we conduct DNSs to investigate the energy transfer mechanisms and the statistical properties of the velocity fluctuations in a  stirred three-dimensional symmetric binary fluid mixture.  Our main findings are: $(i)$  External stirring arrests phase-separation (coarsening); $(ii)$ Our scale-by-scale analysis reveals that interfaces provide an alternate route for energy transfer and dissipation; $(iii)$ For same Reynolds number, a single-component fluid and a binary-fluid mixture have same small-scale statistics. 

The rest of the paper is organised as follows.  We present the equations and the numerical method that we use in Sec.~\ref{sec:eq}.  In Sec~\ref{sec:eeb}-\ref{sec:eflx}  we derive the equations for the total energy, the scale-by-scale kinetic energy budget, and present our numerical findings.  In Sec.~\ref{sec:ss} we investigate the small-scale statistics of the vorticity magnitude and the energy dissipation rate. We conclude in Section~\ref{sec:ccl}.

\section{\label{sec:eq} Equations and Direct Numerical Simulation (DNS)}
We model symmetric binary-fluid mixture by using the Navier-Stokes-Cahn-Hilliard (NSCH) or Model-H \citep{cah68,hoh77} equations
\begin{eqnarray}
\rho (\partial_t + \bm u \cdot \nabla) {\bm u} &=& \eta \nabla^2  {\bm u} -\nabla P   + \Lambda \phi \nabla \nabla^2 \phi + f,  \label{ch:eq1} \\
(\partial_t + {\bm u} \cdot \nabla) {\phi} &=& M \nabla^2 \mu, \label{ch:eq2} \\
\nonumber
\mu&\equiv& \Lambda \left(\frac{-\phi + \phi^3 }{\xi^2}-  \nabla^2 \phi \right),~{\rm{and}} \\
\nonumber
\nabla \cdot {\bm u} &=&0. 
\nonumber
\end{eqnarray}
Here ${\phi}, {\bm u}$, and $P$ are the Cahn-Hilliard order-parameter,  the velocity, and the pressure field at position ${\bm x}$ and time $t$, $\eta$ is the dynamic viscosity, $\rho$ is the density, the kinematic viscosity $\nu \equiv \eta/\rho$,  $M$ is the mobility,  $\xi$ controls the width of the interface between the two phases, $\Lambda$ is the mixing energy density, the order-parameter diffusivity $\kappa= M \Lambda/\xi^2$, the surface tension $\sigma= 2 \sqrt{2}/3 (\Lambda/\xi)$, and ${\bm f}$ is the external forcing that generates turbulence. The order parameter takes positive values in one phase and negative in the other. For simplicity, we assume the density ($\rho=1$), the viscosity $\eta$, and the mobility $M$ to be independent of $\phi$ and same for the two phases  \footnote{In the LB simulations of \cite{per14}, mobility depends on $\phi$.}. The local vorticity ${\bm \omega}\equiv \nabla \times {\bm u}$. Note that the pressure $P$ also contains contributions due to longitudinal terms in $\phi$.  We use a cubic domain with each side of length ${\cal L}=2\pi$ and discretize it with  $N^3$ collocation points. We employ periodic boundary conditions. Eqs.~\eqref{ch:eq1} and \eqref{ch:eq2} are numerically integrated using a pseudo-spectral method with $1/2$-dealiasing and time marching is done using an exponential Adams-Bashforth scheme \citep{cox02}. A large-scale forcing $\hat{\bm f}_{\bm k}= f_0 \hat{\bm u}_k/\sum_{k=1,2}|{\bm u}_k|^2$, where the caret indicates Fourier transform, with $|{\bm k}|\leq 2$ ensures constant energy injection rate $\epsilon_{inj}$. 

\section{\label{sec:res}Results}
The energy injection based Reynolds number $\Rey \equiv  \sqrt{\epsilon_{inj}^{1/3} {\cal L}^{4/3}/\nu}$ and Weber number $\We \equiv  \rho \epsilon_{inj}^{2/3} {\cal L}^{5/3}/\sigma$ characterise the turbulence intensity of the flow. Table~\ref{tab1} summarises the parameters that we use.  All the simulations were time-integrated up to $t=100 T_{\cal L}$ ($T_{\cal L}\equiv \epsilon_{inj}^{1/3} {\cal L}^{-2/3}$). We collect the statistics after a steady-state is attained $(t > 40 T_{\cal L})$.

\begin{table}
\begin{center}
   \begin{tabular}{lccccccccccc}
    $ $ &$N$ & $\nu$ & $\xi (\times 10^{-2})$ & $\kappa$ & $\sigma (\times 10^{-1})$ & $\Rey$ & $\We$ & $\epsilon_{inj}$ & $\epsilon_\nu (\times 10^{-1})$ & $\epsilon_\mu (\times 10^{-1})$    \\
    {\tt NS1}  & $256$ & $5 \cdot 10^{-3}$ & $-$ & $-$ & $-$ &  $65$  &  $-$ & $0.05$ & $0.05$ & $-$  \\
    {\tt SP11}  & $256$ & $5 \cdot 10^{-3}$ & $1.88$ & $10^{-2}$ & $2.51$ &  $65$  &  $11.6$ & $0.05$ & $0.33$ & $0.18$  \\
     {\tt SP12}  & $256$ & $5 \cdot 10^{-3}$ & $1.88$ & $10^{-2}$ & $0.63$ &  $65$  &  $46.4$ & $0.05$ & $0.23$ & $0.28$  \\
      {\tt SP13}  & $256$ & $5 \cdot 10^{-3}$ & $1.88$ & $10^{-2}$ & $0.25$ &  $65$  &  $116.2$ & $0.05$ & $0.23$ & $0.27$  \\
     {\tt NS2}  & $512$ & $5 \cdot 10^{-3}$ & $-$ & $-$ & $-$ &  $96$  &  $-$ & $0.5$ & $0.5$ & $-$  \\
      {\tt SP21}  & $512$ & $5 \cdot 10^{-3}$ & $3.76$ & $10^{-2}$ & $5.02$ &  $96$  &  $26.9$ & $0.5$ & $3.10$ & $1.90$  \\
  {\tt SP22}  & $512$ & $5 \cdot 10^{-3}$ & $3.76$ & $10^{-2}$ & $3.76$ &  $96$  &  $35.9$ & $0.5$ & $2.90$ & $2.15$  \\
  {\tt SP23}  & $512$ & $5 \cdot 10^{-3}$ & $3.76$ & $10^{-2}$ & $2.50$ &  $96$  &  $53.9$ & $0.5$ & $2.67$ & $2.35$  \\
  {\tt SP24}  & $512$ & $5 \cdot 10^{-3}$ & $3.76$ & $10^{-2}$ & $1.25$ &  $96$  &  $107.9$ & $0.5$ & $2.36$ & $2.55$  \\
\end{tabular}
\caption{\label{tab1} \small Parameters $N, \nu, \xi, \kappa, \sigma, Re, We, \epsilon_{inj}, \epsilon_\nu, \epsilon_\mu$ for our binary-fluid DNS  {\tt SP11-3} and {\tt SP21-4}.  The ${\tt NS}$ runs are the DNS studies conducted for single-component Navier-Stokes fluid with the same $\nu$ and $\epsilon_{inj}$ as the  binary-fluid.}
\end{center}
\end{table}

\begin{figure}
\begin{center}
\includegraphics[width=0.45\linewidth]{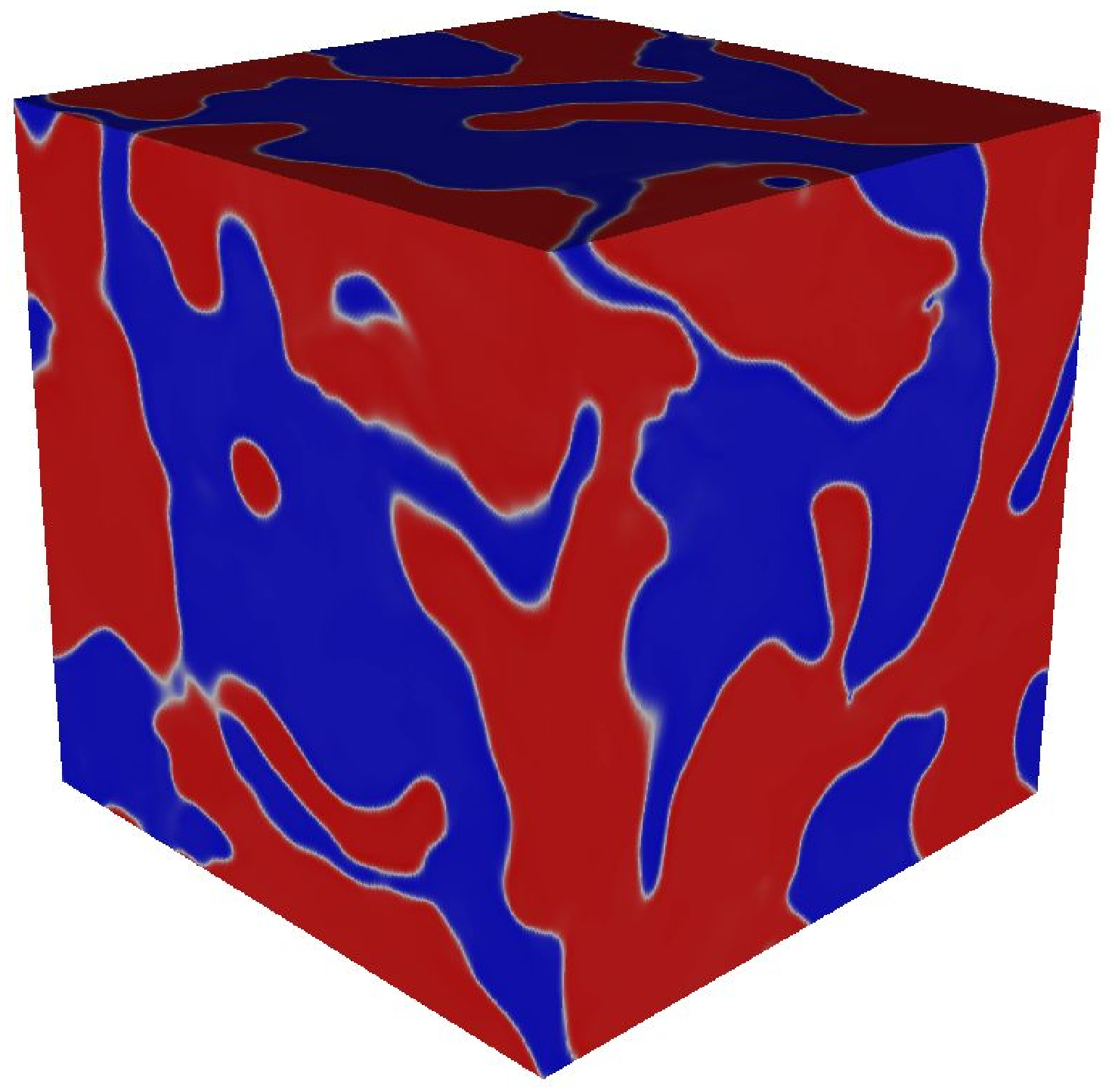}
\includegraphics[width=0.45\linewidth]{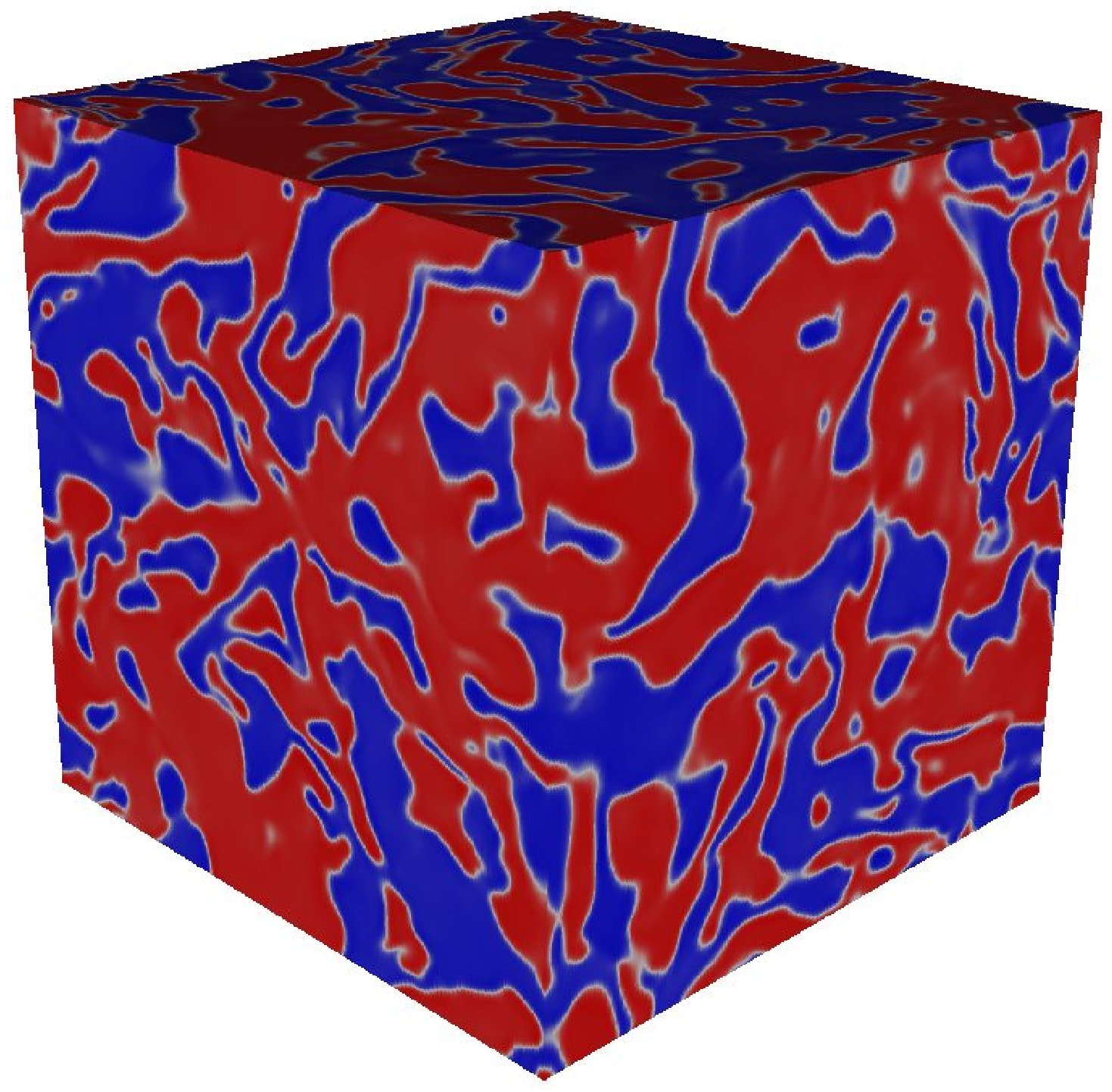} 
\end{center}
\caption{\label{fig:pvor}  Pseudocolor plot of the order parameter field $\phi$ for $\Rey=96$, $\We=26.9$ (left) and  for $\Rey=96, \We=107.9$ (right). Regions with $\phi=1 (-1)$ are shown in red (blue). Average domain size decreases with increasing $\We$. } 
\end{figure}

\subsection{\label{sec:eeb} Domain size and Energy balance}
\begin{figure}
\centerline{
\includegraphics[width=0.45\linewidth]{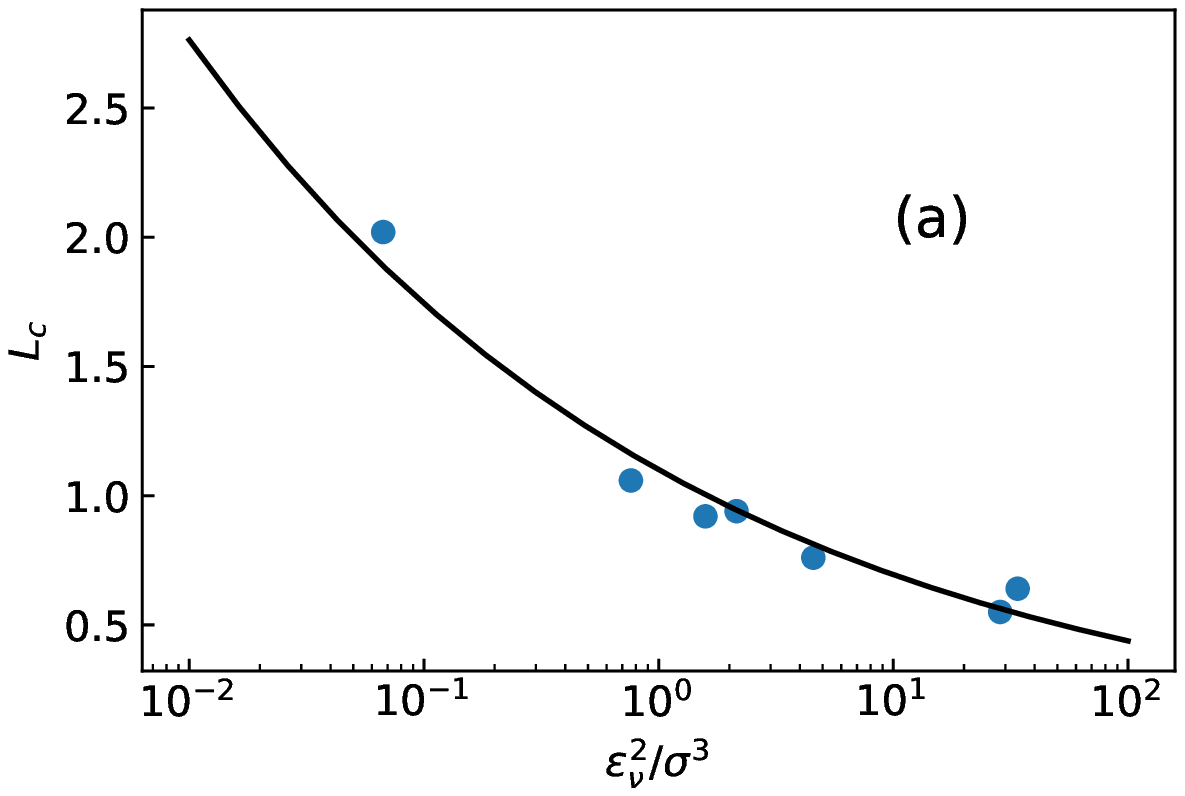}
\includegraphics[width=0.45\linewidth]{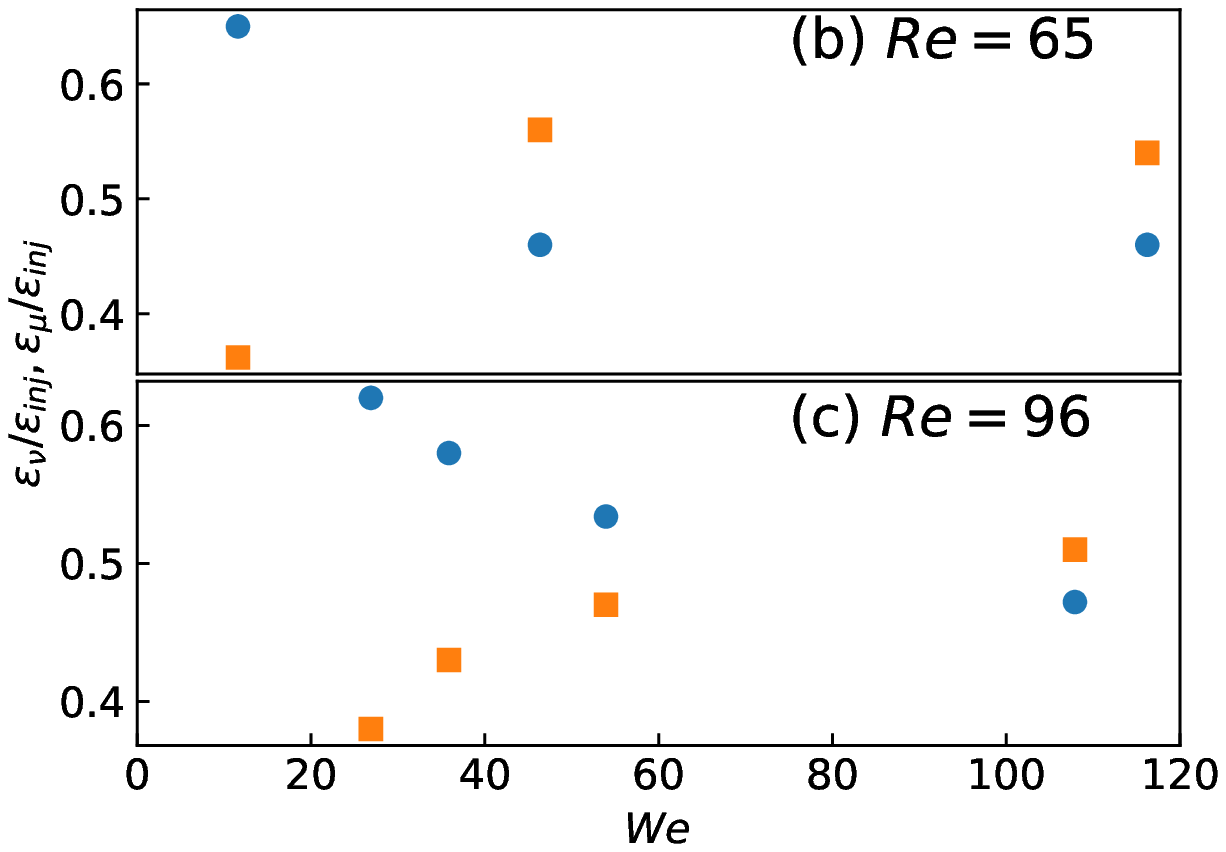}}
\caption{\label{fig:fig1} (a) Semilog plot of the domain scale $L_c$ versus $\epsilon_\nu^2/\sigma^3$ (blue dots) and its comparison with the Hinze prediction for the domain scale  [$D \sim (\epsilon_\nu^2/\sigma^3)^{-1/5}$, black line]. (b,c) $\epsilon_\nu/\epsilon_{inj}$ (blue circle) and $\epsilon_\mu/\epsilon_{inj}$ (orange square) with varying Weber number $\We$ for $\Rey=65$ (runs {\tt SP11-13}) and $\Rey=96$ (runs {\tt SP21-24}).}
\end{figure}

 In Fig.~\ref{fig:pvor}  we plot the representative snapshots of the steady-state order-parameter field $\phi$ for our runs  ${\tt SP21}$ $( \Rey=96, \We=26.9)$and ${\tt SP24}$ $(\Rey=96, \We=107.9)$. We observe that the domain size in the emulsion decreases with increasing $\We$. From $\phi$, we estimate the average domain size as
\begin{equation}
L_c=2\pi \left \langle \frac{\sum_{\bm k}   |\hat{\phi}_{\bm k}|^2}{\sum_{\bm k}  k |\hat{\phi}_{\bm k}|^2}\right \rangle, 
\end{equation}
where $k=\sqrt{{\bm k} \cdot {\bm k}}$ and angular brackets denote time-averaging in the statistically steady-state. In \subfig{fig:fig1}{a} we present a consolidated plot of $L_c$  versus $\epsilon_\nu^2/\sigma^3$ for different value of $\Rey$ and $\We$ (see Table~\ref{tab1}). We find the data to be in good agreement with Hinze's prediction for the average domain size in an emulsion $D\sim (\rho \epsilon_\nu^2/\sigma^3)^{-1/5}$.

We next investigate how much of the energy injected $\epsilon_{inj}$ is dissipated by the viscosity $\epsilon_\nu$ and how much is the interfacial contribution $\epsilon_\mu$. Using  \Eq{ch:eq1} and \Eq{ch:eq2}, we obtain the following energy balance  equation: 
\begin{equation}
\begin{split}
\partial_t \underbrace{\int \frac{|\bm u|^2}{2} d{\bm x}}_{\cal K} + &  \partial_t \underbrace{\int  \frac{\Lambda}{\xi^2}\left(-\frac{\phi^2}{2} + \frac{\phi^4}{4} + \frac{\xi^2}{2}  |\nabla \phi|^2 \right)  d{\bm x}}_{\cal G}  \\ &\quad = -\underbrace{\nu \int |\nabla u|^2  d{\bm x}}_{\epsilon_\nu}  - \underbrace{M  \int |\nabla \mu|^2 d{\bm x}}_{\epsilon_\mu} +  \underbrace{\int {\bm f} \cdot {\bm u} d{\bm x}}_{\epsilon_{inj}}.
\end{split}
\end{equation}
Here ${\cal K}$ is the kinetic energy, ${\cal G}$ is the free energy of mixing, $\epsilon_\nu$ is the viscous energy dissipation, $\epsilon_\mu$ is the dissipation because of the chemical potential (interfacial contribution), and $\epsilon_{inj}$ is the energy injected because of the external forcing.  In the statistically steady-state,  $\langle \epsilon_{inj} \rangle \approx \langle \epsilon_\nu \rangle + \langle \epsilon_\mu \rangle$. In  \subfig{fig:fig1}{b,c}  we show that for small $\We$, the viscosity is the primary dissipation mechanism $\langle \epsilon_\nu \rangle \gg  \langle \epsilon_\mu \rangle$ whereas, for large $\We$  average domain size reduces thereby making the interfacial contribution more dominant $\langle \epsilon_\mu \rangle > \langle \epsilon_\nu \rangle$.  

\subsection{\label{sec:espec}Energy spectrum}

The steady-state  energy spectrum is defined as \citep{vin91}
\begin{equation}
E(k)= \frac{1}{2} \left \langle \sum_{k-1/2<k^\prime \leq k+1/2} |\hat{\bm u}_{\bm k^\prime}|^2 \right \rangle. 
\end{equation}

\begin{figure*}
\centerline{
\includegraphics[width=0.48\linewidth]{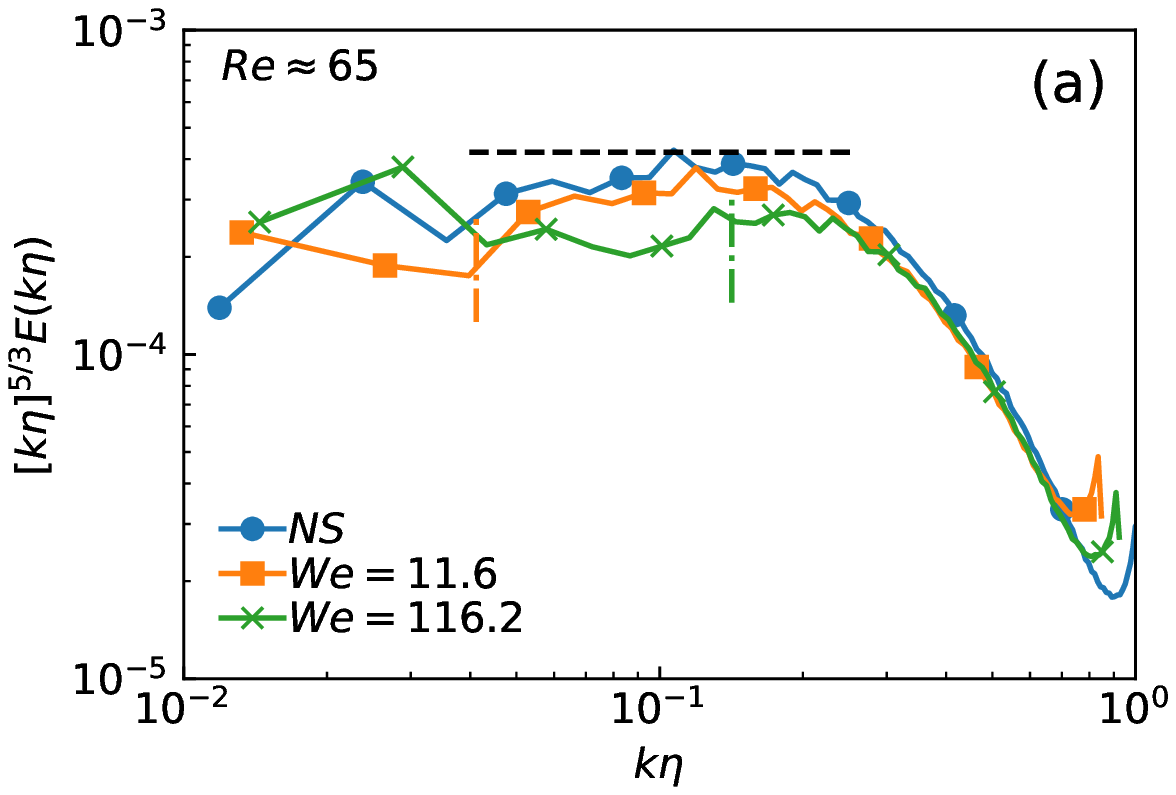}
\includegraphics[width=0.48\linewidth]{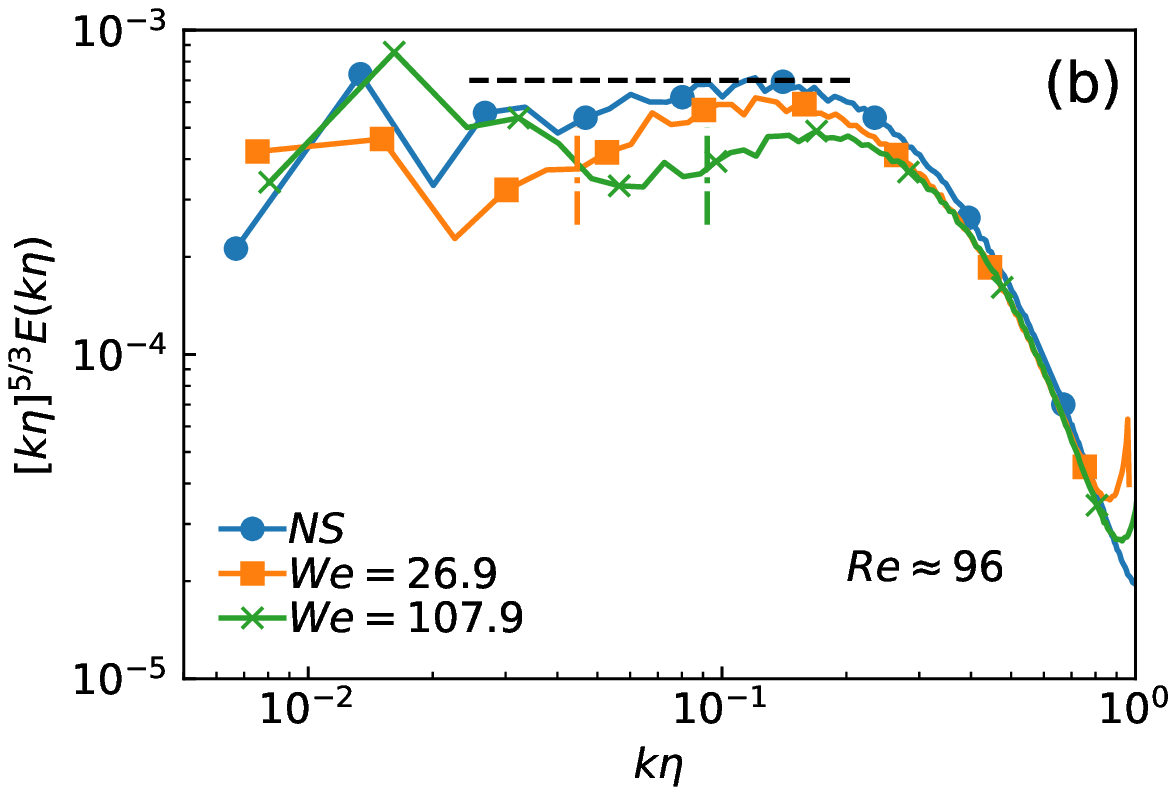}
}
\caption{\label{fig:esp}  Energy spectrum  for (a) $\Rey=65$ and  (b) $\Rey=96$ for different values of $\We$. The horizontal dashed line indicate Kolmogorov scaling and the vertical dash-dot indicate the wave-number corresponding to the average domain size $2 \pi/L_c$ in the emulsion.}
\end{figure*}

For high Reynolds number single-component flows, $E(k)\sim k^{-5/3}$ in the inertial range \citep{kol41,Fri96}. The energy spectrum obtained from our DNS run ${\tt NS1}$ and ${\tt NS2}$ shows 
half a decade of scaling for $\Rey=65$ and nearly a decade of scaling for $\Rey=96$ [see \Fig{fig:esp}].  For the same Reynolds number, the inertial range energy spectrum is suppressed for binary-fluid mixtures for $k < 2\pi/L_c$. For scales smaller than the average domain size of the emulsion $k>2\pi/L_c$, the energy spectrum follows its single-component counterpart.  Our high-$\Rey=96$, high-$\We=107.9$ run indicate that Kolmogorov scaling is recovered for $k\ll 2\pi/L_c$. Numerical simulations with large scale-separation and much higher spatial resolutions are required to investigate the recovery of Kolmogorov scaling for $2\pi/\ell_{inj} \gg k \gg 2 \pi/L_c$.  

\REM{For high Reynolds number single-component fluid turbulence $E(k)\sim  k^{-5/3}$ \citep{kol41,Fri96} in the inertial range. In \Fig{fig:esp} we plot the compensated energy spectrum $k^{5/3} E(k)$ showing a half a decade of scaling for $\Rey=65$ and nearly a decade of scaling for $\Rey=96$. In binary-fluid mixtures this scaling is not observed for wave-numbers $k< 2\pi/L_c$ and the energy content of these wave-numbers is reduced in comparison to the pure fluid case.  For $k>2\pi/L_c$ we find indications that $E(k)$ for binary-fluid is close to its pure fluid counterpart. However, much higher resolution simulations are required to verify if the Kolmogorov scaling survives for $k> 2\pi/L_c$.}

\subsection{\label{sec:eflx}Scale-by-scale Kinetic Energy budget}
\label{sec:eflux}
To investigate how the kinetic energy is distributed upto a given length scale $\ell$ (or a corresponding wave-number $k=2\pi/\ell$), we now derive the scale-by-scale energy budget equation. By multiplying the Fourier transformed \Eq{ch:eq1} with $\hat{\bm u}_{-\bm k}$,  summing up contributions upto wave-number $k$, and then averaging over statistically-steady state  we obtain 
\begin{equation}
\Pi(k) =    - {\cal D}(k) - \Pi^\phi(k) + {\cal F}(k).
\end{equation}
Here $\Pi(k) \equiv \langle \Re[\sum_{|{\bm m}|\leq k}\hat{{\bm u}}_{\bm m} \cdot \widehat{(\bm{u} \cdot \nabla {\bm u})}_{-\bm m}] \rangle$ is the energy flux, $\Pi^{\phi}\equiv \Lambda \langle \Re[ \sum_{|{\bm m}|\leq k} \hat{{\bm u}}_{\bm m} \cdot \widehat{(\phi \nabla \nabla^2 \phi)}_{-\bm m}] \rangle$ is the cumulative flux of $(\nabla \phi)^2$, ${\cal E}(k) \equiv \sum_{m\leq k} E(m)$ is the cumulative energy upto $k$,  ${\cal D}(k)\equiv \nu \sum_{m\leq k}  m^2 E(m)$ is the cumulative dissipation upto wave-number $k$, and ${\cal F}(k)\equiv \langle  \sum_{|{\bm m}| \leq k} \Re [ {\bm u}_{-\bm m} {\bm f}_{\bm m}] \rangle$ is the cumulative energy injected.  In statistically steady state the left hand size is zero and $\Pi(k) = {\cal D}(k) - \Pi^\phi(k) + {\cal F}(k)$. When $K=2\pi/{\cal L}$,  $\Pi=0, {\cal D}=\epsilon_\nu, \Pi^\phi = \epsilon_\mu,$ and ${\cal F}=\epsilon_{inj}$. 

\begin{figure*}
\begin{center}
\includegraphics[width=0.4\linewidth]{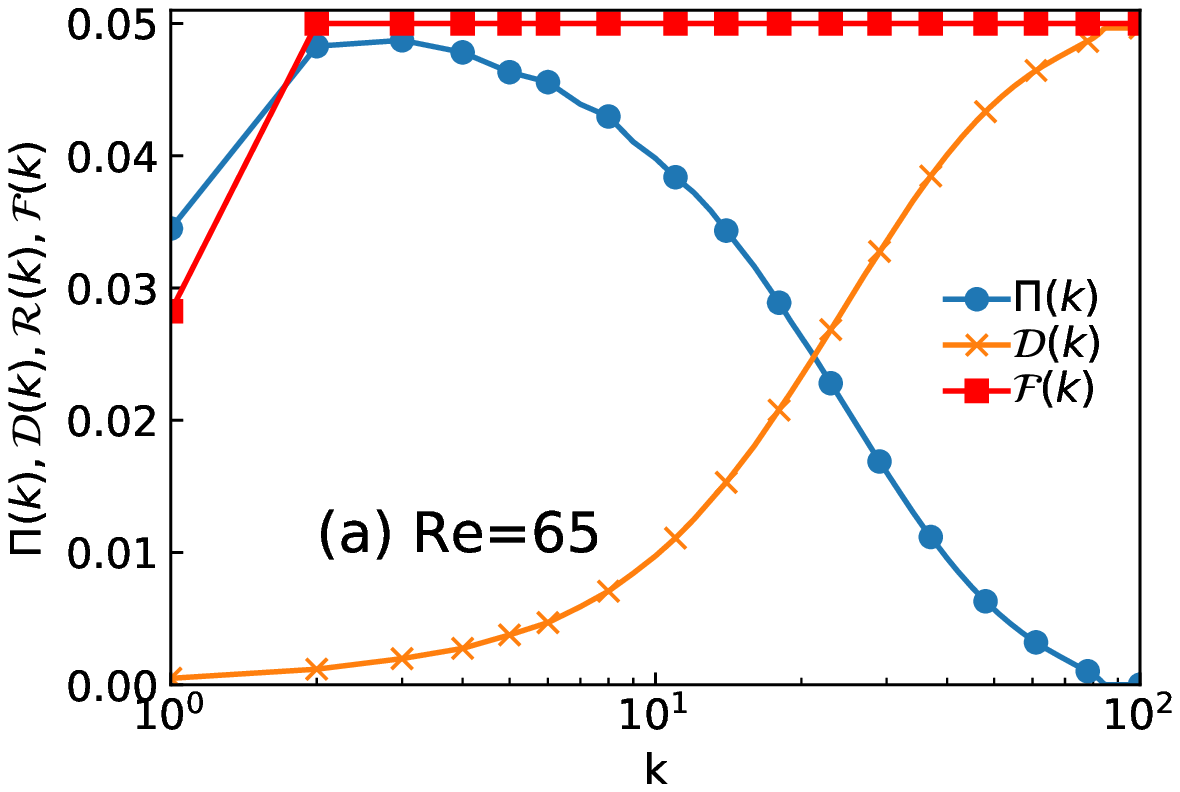}
\includegraphics[width=0.4\linewidth]{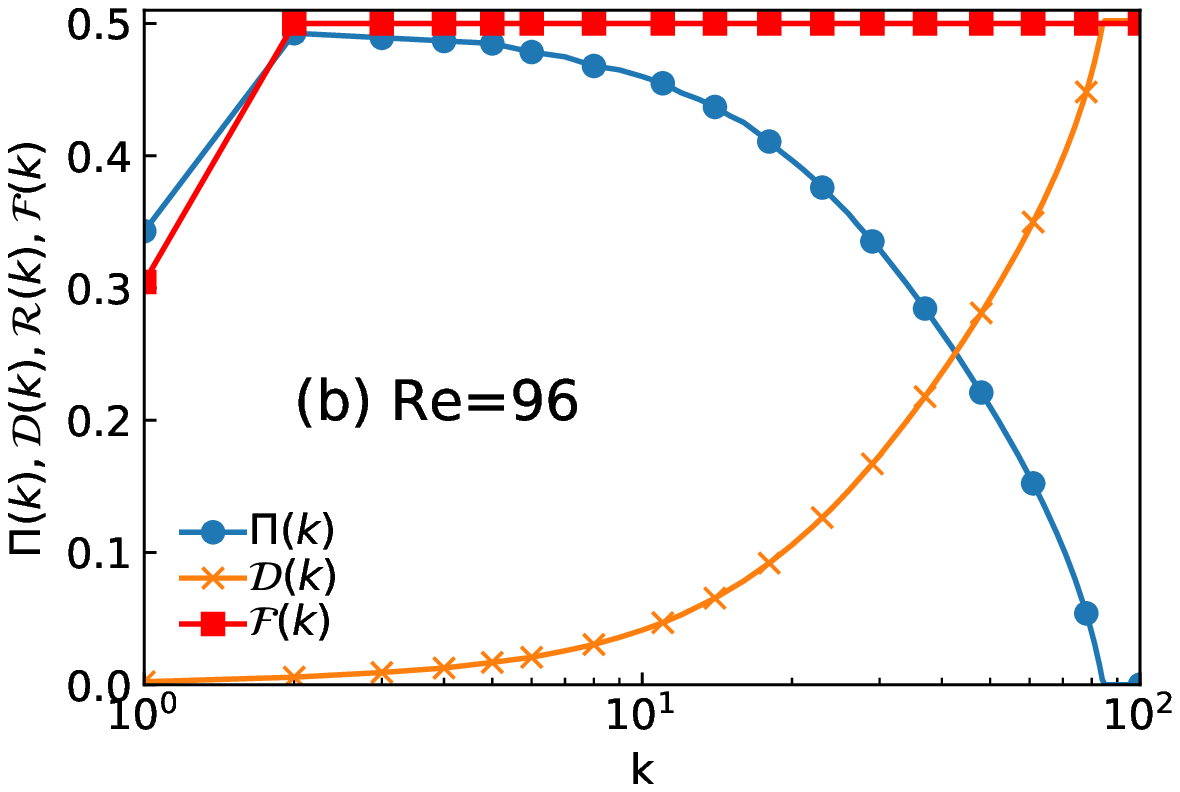}\\
\includegraphics[width=0.4\linewidth]{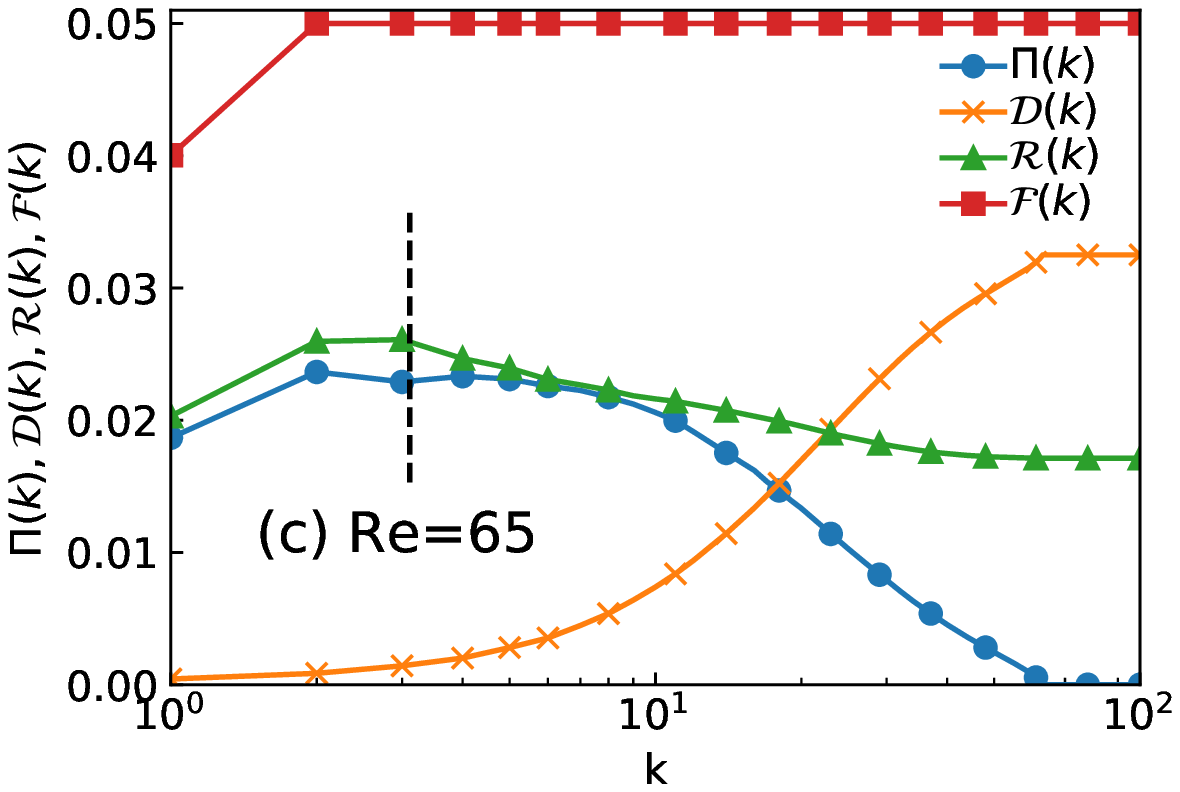}
\includegraphics[width=0.4\linewidth]{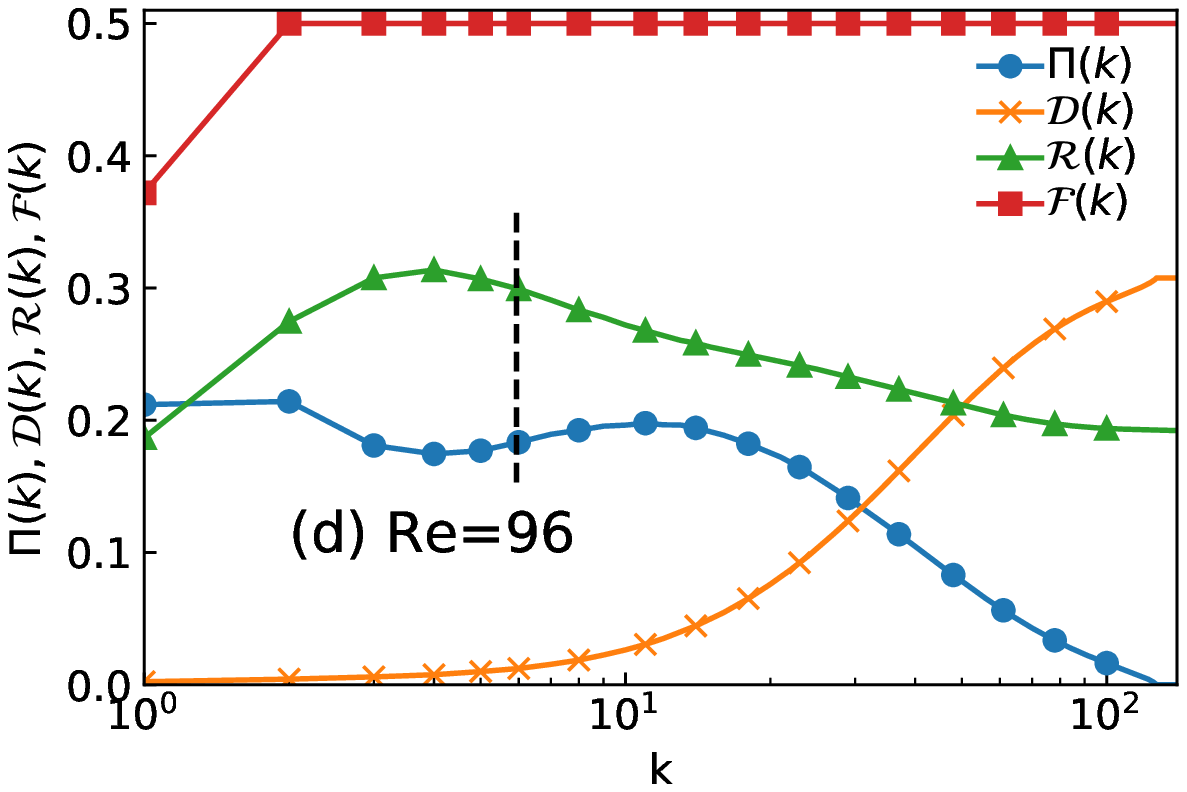}\\
\includegraphics[width=0.4\linewidth]{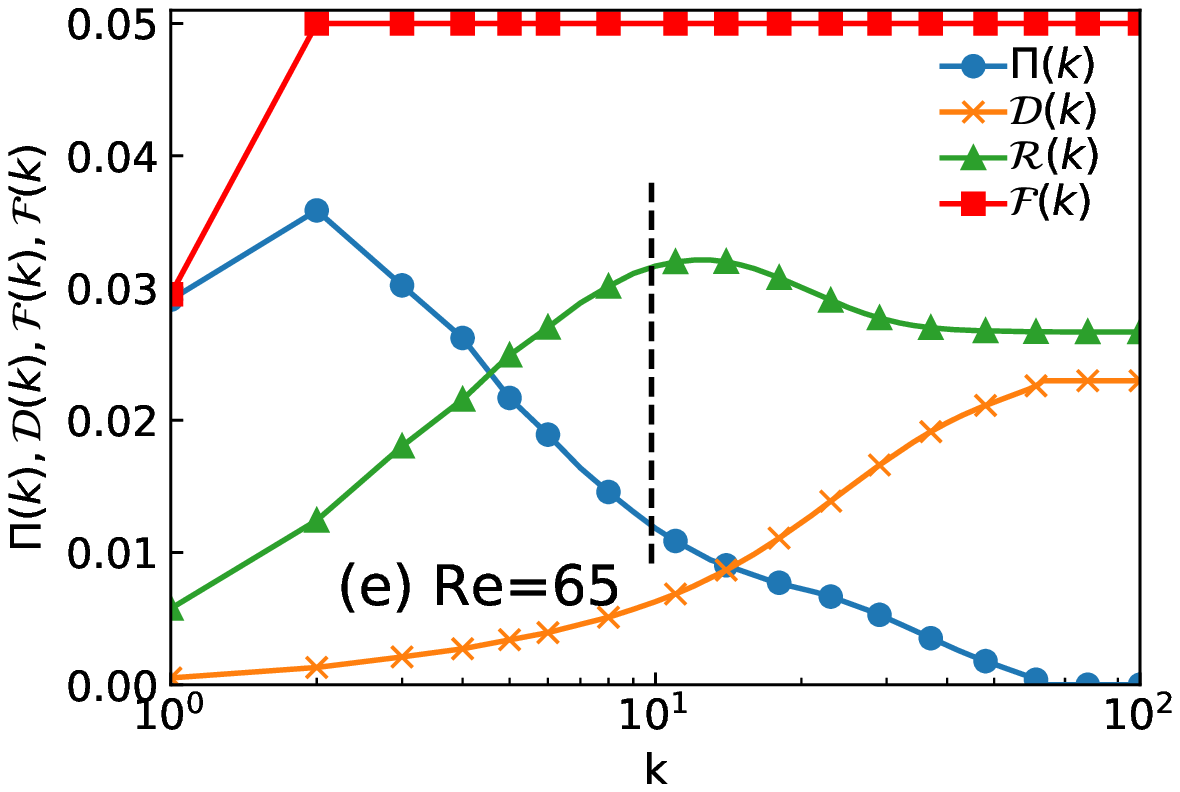}
\includegraphics[width=0.4\linewidth]{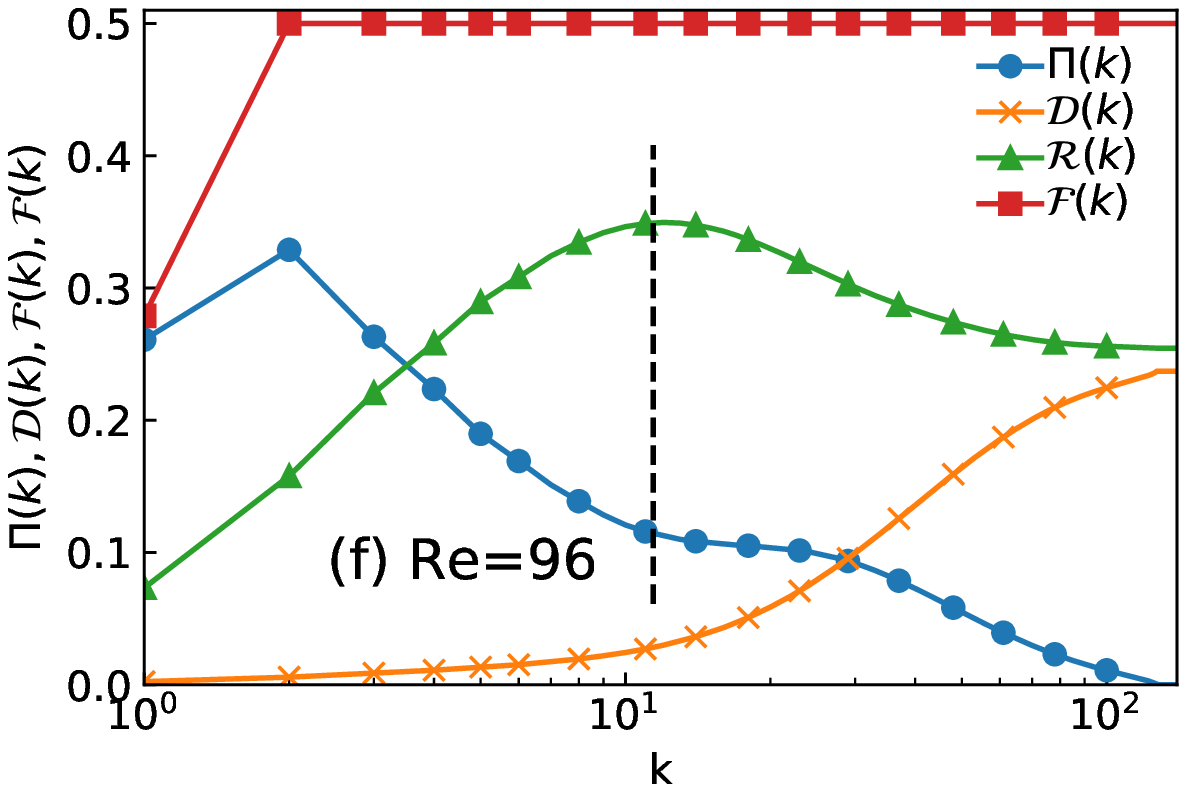}
\end{center}
\caption{\label{fig:flx1}  (Top panel) Energy flux for (a) $\We\Rey=65$ (run {\tt NS1}), (b) $\Rey=96$ (run {\tt NS2}), (c) $We\approx 11.6, \Rey \approx 65$ (run {\tt SP11}), (d) $We \approx 26.9, \Rey \approx 96$ (run {\tt SP21}),  (e) $We \approx 116.2, \Rey \approx 65$ (run {\tt SP13}), and (f) $We \approx 107.9, \Rey \approx 96$ (run {\tt SP24}). The dashed vertical line indicate the wave-number corresponding to the average domain size $2 \pi/L_c$.}
\end{figure*}

The plot in \subfig{fig:flx1}{a,b} shows the scale-by-scale energy budget for $\Rey=65$ and $\Rey=96$ for a single-component fluid. The energy is injected at large-scales by forcing ${\cal F}(k)$  and is dissipated by viscosity at small-scales. The Navier-Stokes nonlinearity transfers the kinetic energy in the inertial range while keeping its flux $\Pi(k)$ constant.  For our high $\Rey=96$ run ${\tt NS2}$, we observe a nearly constant $\Pi(k) \sim \epsilon_{inj}$ for $2\leq k \leq 10$ which manifests as  an extended inertial range in the energy spectrum [see \subfig{fig:esp}{b}]. 

We now show that  for the binary fluid case, the presence of emulsion domains dramatically alter the energy transfer mechanism. Firstly, for small $\We$, we note that the viscous dissipation $\epsilon_\nu \equiv {\cal D}(2\pi/{\cal L})$ is larger than the interfacial dissipation $\epsilon_\mu \equiv \Pi^\phi(2\pi/{\cal L})$ [\subfig{fig:flx1}{b,d}] whereas it is the opposite for high $\We$ [\subfig{fig:flx1}{e,f}]. Secondly, the interfacial flux $\Pi^{\phi}(K)$ first increases up to a wave-number corresponding to the arrest scale $k_c \approx 2\pi/L_c$.  After $k>k_c$, $\Pi^\phi(k)$ decreases until for large $k$ where it is equal to $\epsilon_\mu$. This indicates that the interface undulations absorb kinetic energy  $\sim \Pi^{\phi}(k_c)$  from large scales. However only part of it, $\epsilon_\mu$,  is dissipated and the excess energy $\sim \Pi^{\phi}(k_c)-\epsilon_\mu$ is redistributed among wave-numbers $k>k_c$. Finally, the contribution because of the kinetic energy flux $\Pi(K)$ is nearly halved in comparison to its single-component fluid counterpart. For $k<k_c$ we find that $\Pi(k)$ decreases with increasing $k$. For our high $\Rey=96$ runs we find a small range where $\Pi(k) \sim \rm{constant}$ after which it again decreases. This is consistent our earlier observation that for binary-fluid Kolmogorov-scaling in the energy spectrum is recovered for $k>2\pi/L_c$ [see \Fig{fig:esp}] .   Thus presence of interface opens up an additional mechanism to transfer kinetic energy from large-scales to scales smaller than the average size of a domain in the emulsion. 

\section{\label{sec:ss}Small-scale structures}
\begin{figure*}
\begin{center}
\includegraphics[width=0.45\linewidth]{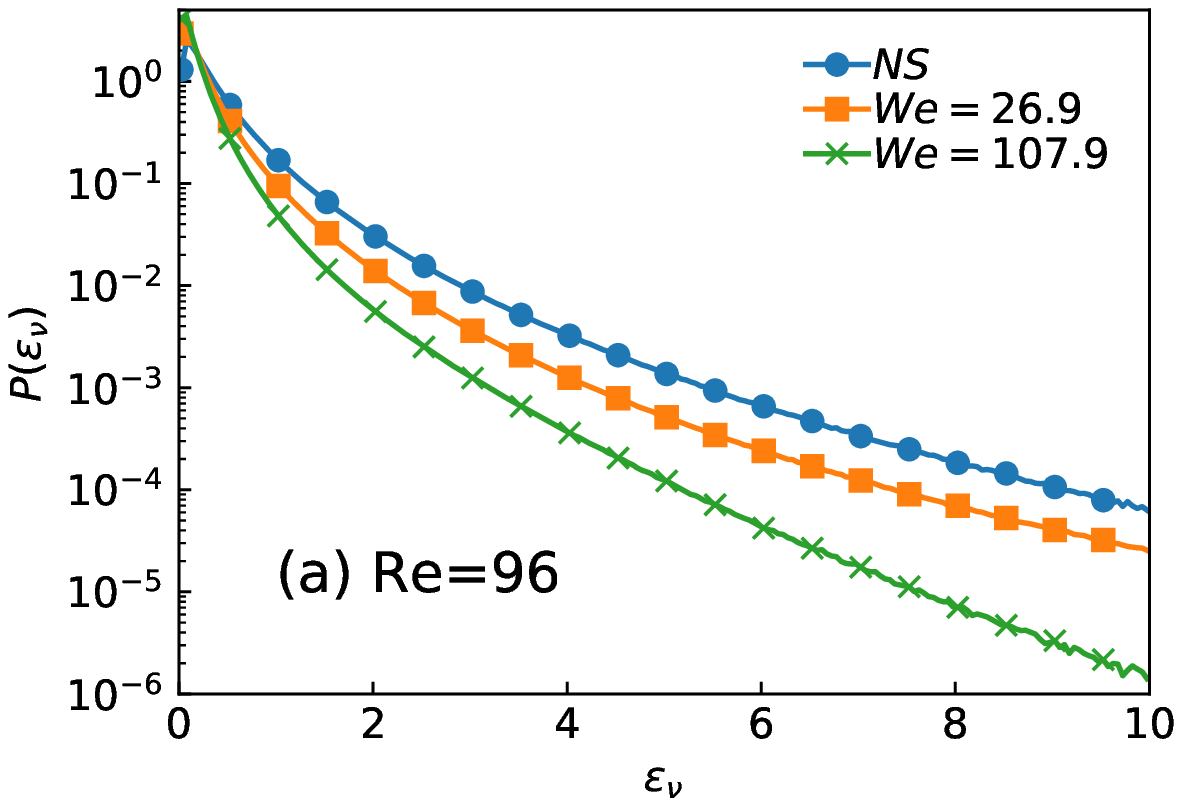}
\includegraphics[width=0.45\linewidth]{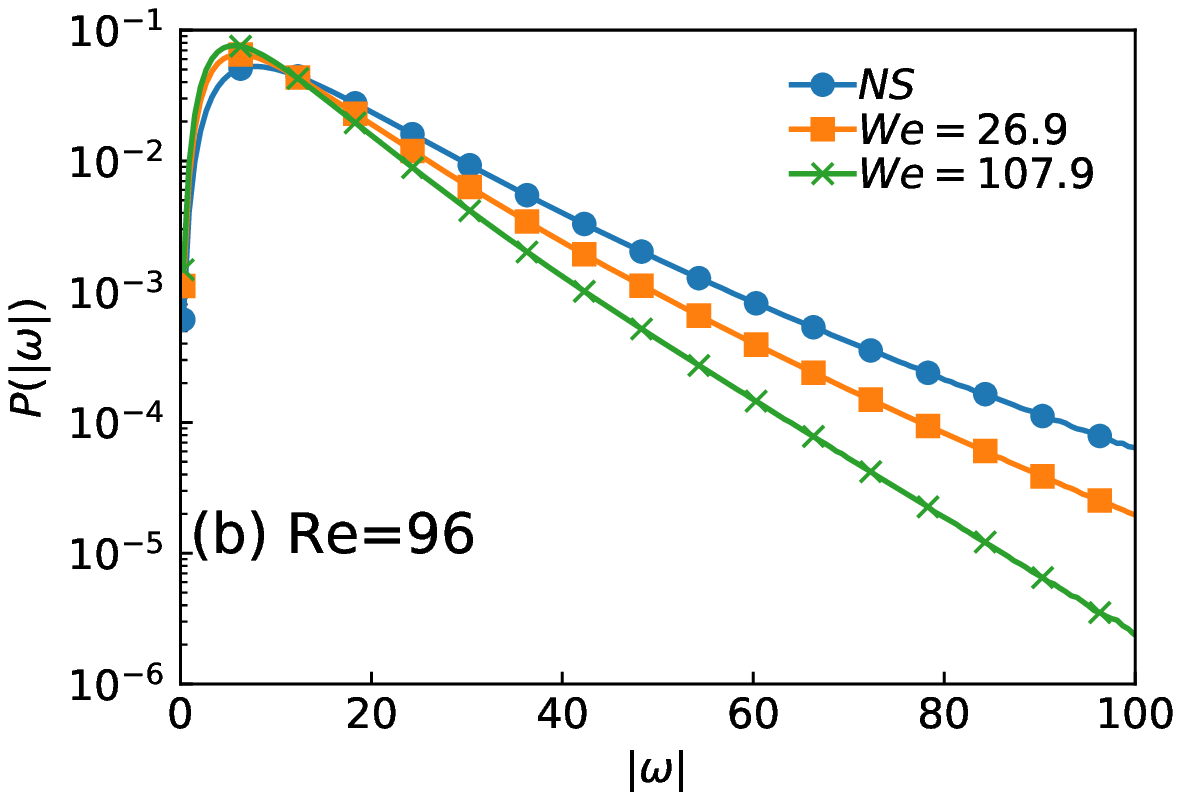}\\
\includegraphics[width=0.45\linewidth]{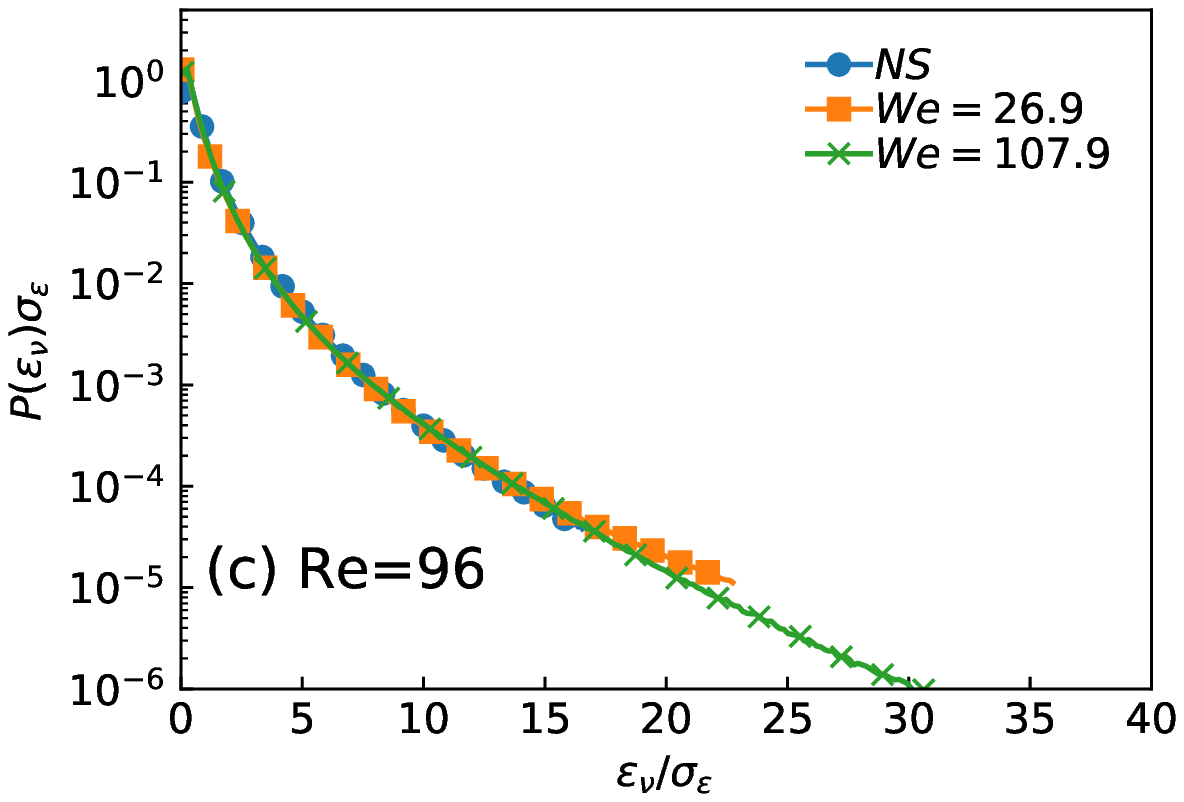}
\includegraphics[width=0.45\linewidth]{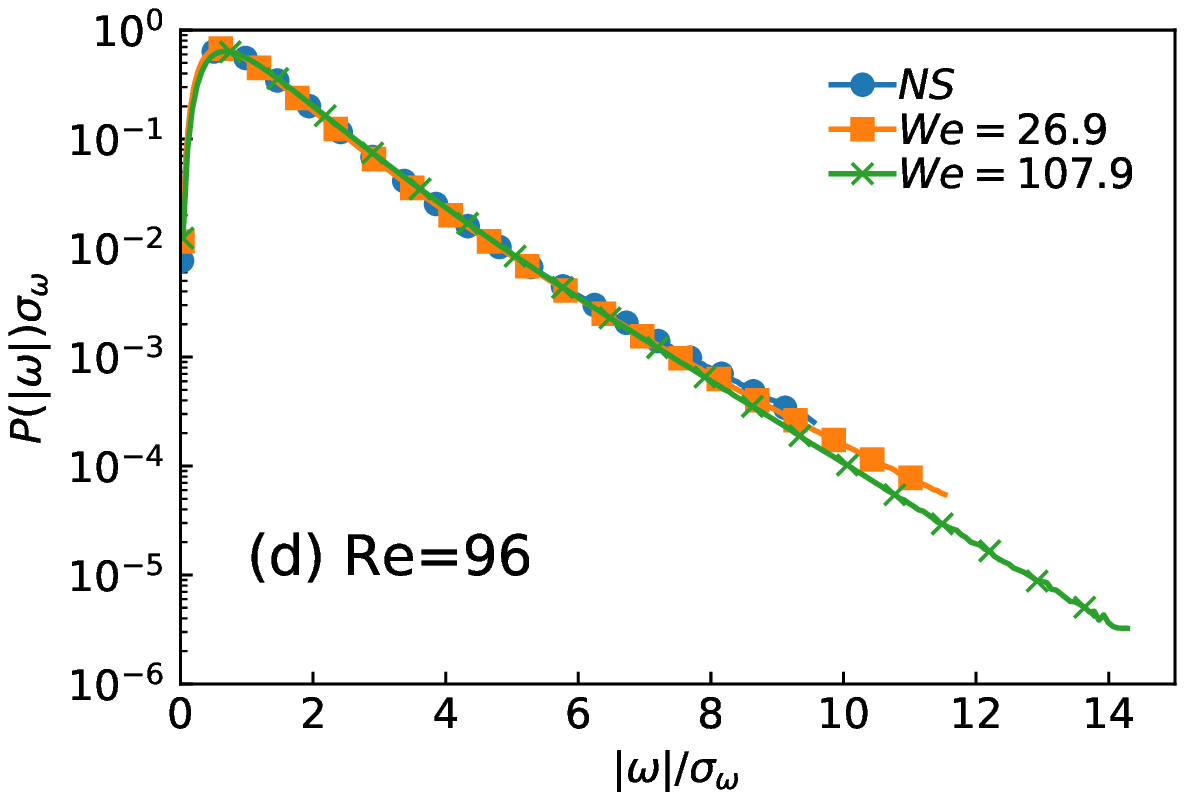}
\end{center}
\caption{\label{fig:eps} (a,b) Probability distribution function (PDF) of the energy dissipation rate $P(\epsilon_\nu)$ versus $\epsilon_\nu$ and the magnitude of the vorticity $P(|\omega|)$ versus $|\omega|$ for different $We$. (c,d) Normalised PDFs $P(\epsilon_\nu) \sigma_\epsilon$  and $P(|\omega|)$ for different $We$. We fix $\Rey=96$ (runs {\tt SP21, SP24}). For comparison we also plot the corresponding PDF for the pure fluid (run {\tt NS2}). }
\end{figure*}

We now investigate whether the different kinetic energy transfer mechanism in binary-fluid turbulence also alter the small-scale structures.   In \subfig{fig:eps}{a,b} we plot the statistics of the local viscous energy-dissipation $\epsilon_{loc}=\sum_{i,j}  (\partial_i u_j + \partial_j u_i)^2/2$ and the magnitude of the vorticity field $|{\bm \omega}| = \sqrt{{\bm \omega}\cdot {\bm \omega}}$ for a binary-fluid mixture and compare it with a single-component fluid at the same $\Rey$.  We observe a reduction in the events with large values of $\epsilon_{loc}$ and $|{\bm \omega}|$ on increasing the $\We$. However, the PDFs overlap when normalised by their standard deviations. Thus the small-scale statistics remains the same as that of a single-component turbulent fluid. 

\begin{figure*}
\begin{center}
\includegraphics[width=0.32\linewidth]{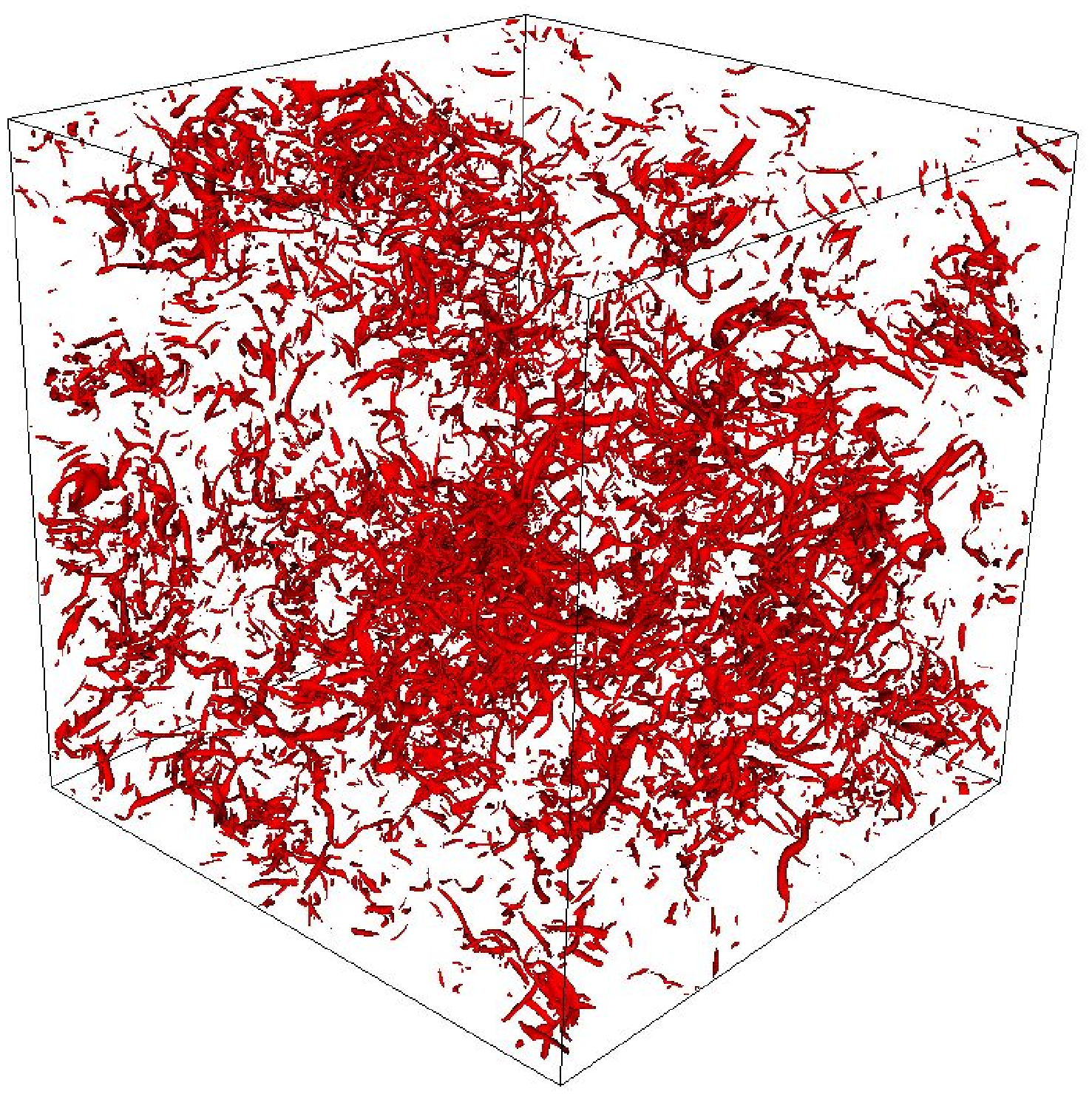}
\includegraphics[width=0.32\linewidth]{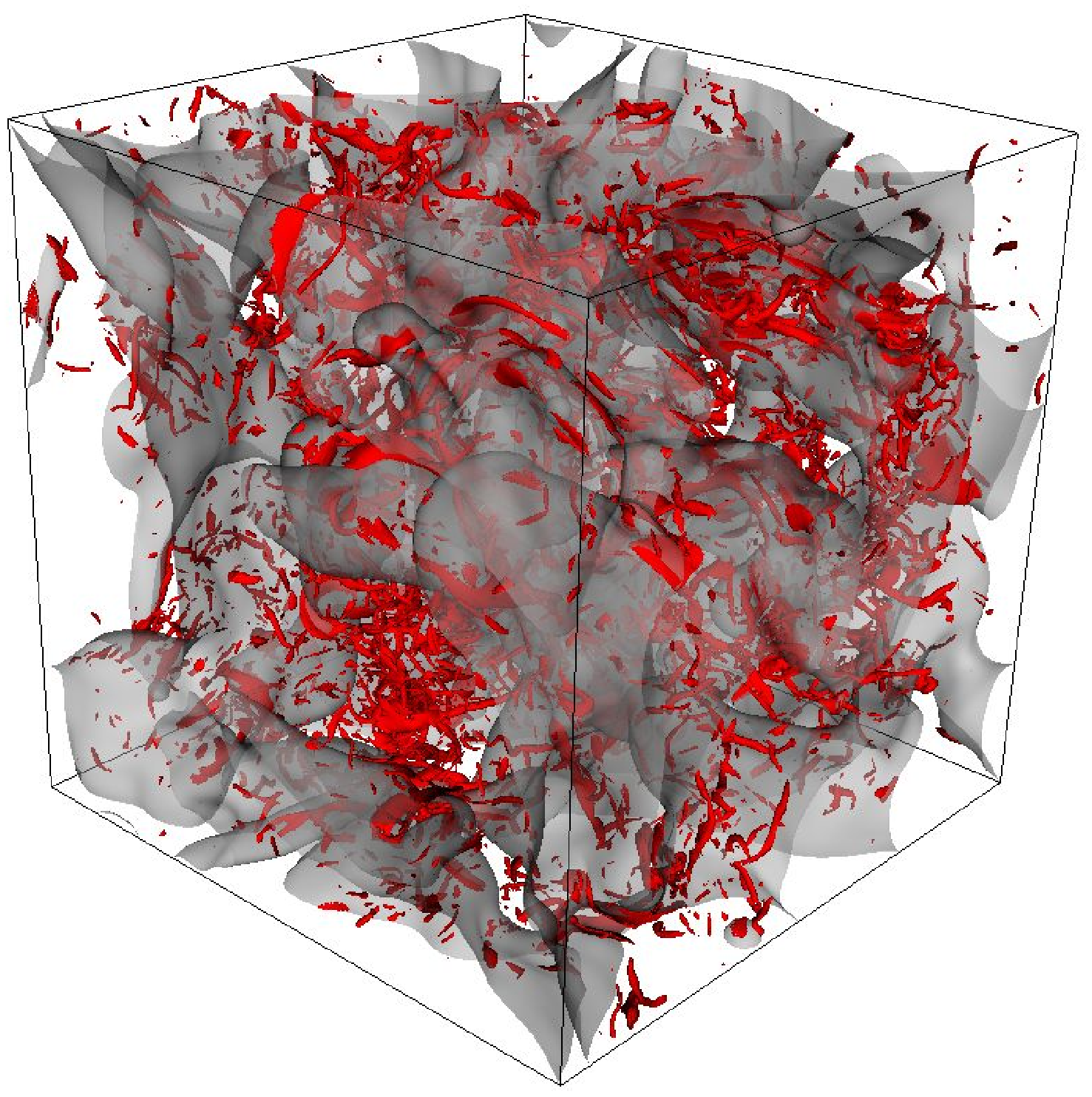}
\includegraphics[width=0.32\linewidth]{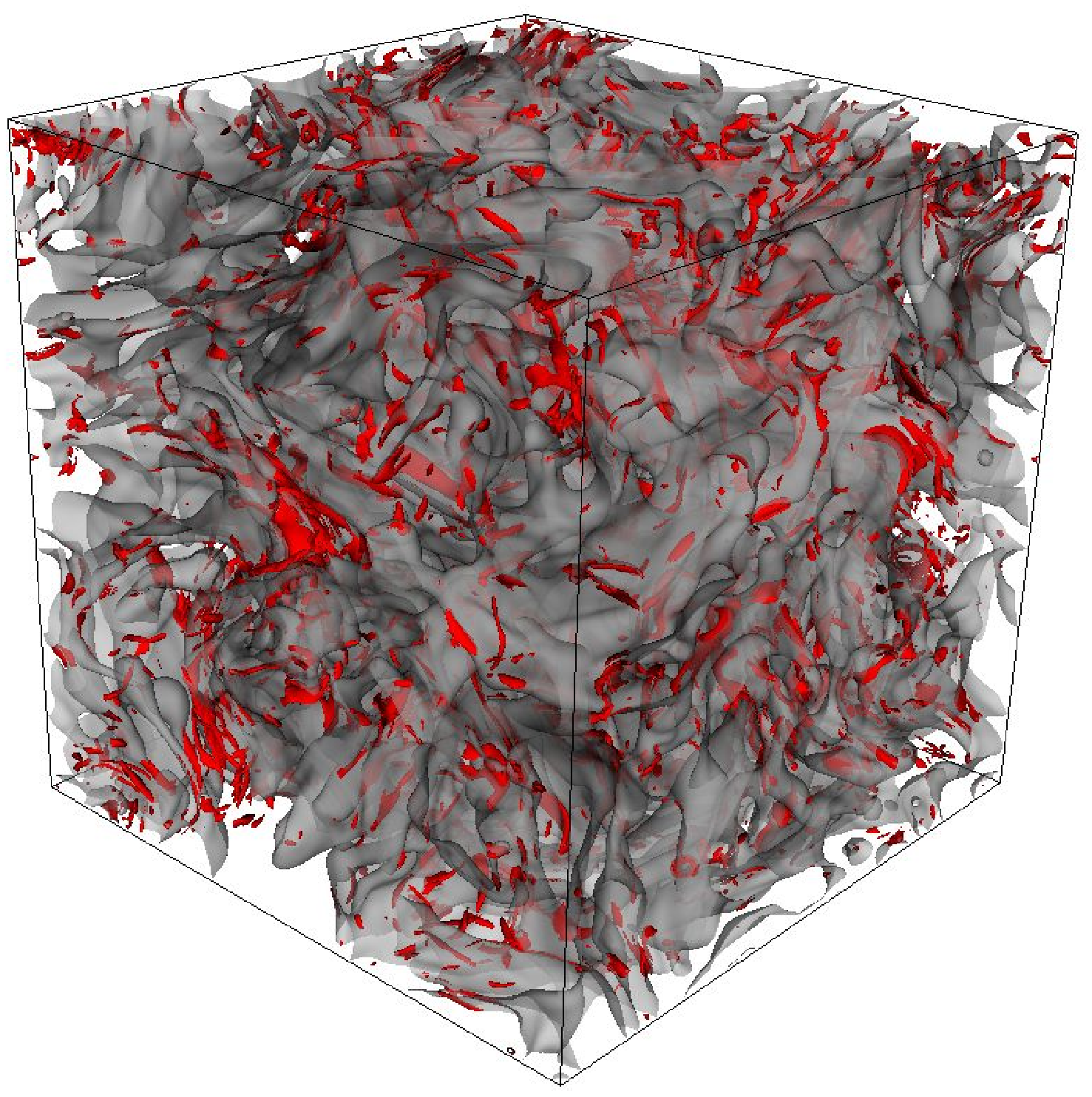}
\end{center}
\caption{\label{fig:isov} Representative steady state iso-contour plots of the vorticity magnitude $|\omega|$ for $|\omega|=\bar{|\omega|} + 6  \sigma_\omega$. (Left) Pure fluid (run ${\tt NS2}$), and a turbulent binary mixture with $\We=26.9$ (center, run ${ \tt SP21}$) and $\We=107.9$ (Right, run ${\tt SP22}$). For the binary mixture plots we also overlay the iso-contour of the order parameter $\phi$ for $\phi=0$ to highlight the interface.}
\end{figure*}

To  investigate the flow structures, in Fig.~\ref{fig:isov} we plot the iso-vorticity contours for a turbulent binary mixture as well as turbulence in pure fluid and overlay the $\phi=0$ contours on them to highlight the emulsion domains. For the single-component fluid, consistent with earlier studies \citep{per09b, ish09}, we observe tubular structures. For the case of binary mixture, as shown earlier, smaller domains (more interfacial area) are formed as we increase $\We$. Also, near the interfacial region vorticity appears to be concentrated. Thus the interface undulations also generate flow structures whose statistics are similar to that of turbulence in a single-component fluid.

\begin{figure*}
\centerline{\includegraphics[width=0.5\linewidth]{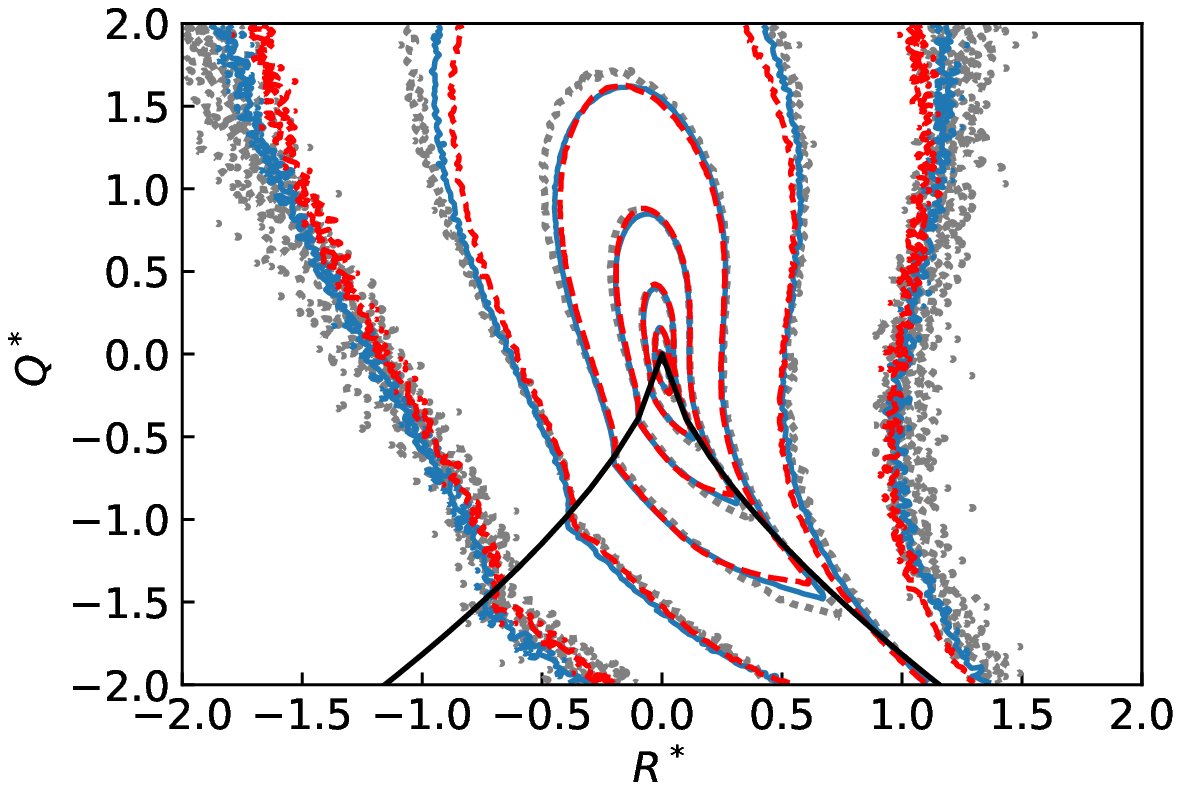}}
\caption{\label{fig:jpdf}Contour plots of the joint PDF $P(R^*,Q^*)$ for pure fluid  turbulence [gray dots, run ${\tt NS2}$], and binary-fluid turbulence with $\We=26.9$ [blue line, run ${\tt SP21}$], $\We=107.9$ [red dashed line, run ${\tt SP24}$]. Here $R^*= \nu^{3/2} R/\epsilon_\nu^{3/2}$ and $Q^*= \nu Q/\epsilon_\nu$. The black curve is the $D=0$ line that demarcates regions with vortical $D>0$ or strain $D<0$ dominated flow structures.}
\end{figure*}

The local flow structures as observed by a Lagrangian fluid parcel in a turbulent flow can be quantified by the invariants $R\equiv -\Tr(A^2)/2$ and $Q\equiv -\Tr(A^3)/3$  of the velocity gradient tensor  $A\equiv \nabla {\bm u}$ \citep{per+cho87,can92}.  The joint PDF $P(R,Q)$ provides an understanding of the typical structures encountered in a turbulent flow. The discriminant  $D\equiv 27 Q^3/4 + R^2$ demarcates  the regions dominated by vortices ($D>0$) with those that are dominated by strain ($D<0$). To quantify whether the structures formed in pure fluid turbulence are similar or different than a turbulent binary mixture,  in \Fig{fig:jpdf} we plot  the joint PDF $P(R^*,Q^*)$ for the pure fluid turbulence as well as binary fluid turbulence with ${\Rey=96}$ and ${\We=26.9,107.9}$. Although the qualitative features remain the same we observe a small but systematic shrinkage  of the iso-contour lines with increasing $\We$.

\section{\label{sec:ccl}Conclusions}
We investigated turbulence in a stirred phase-separated symmetric binary-fluid mixture. Our study reveals that external stirring leads to the formation of a statistically steady emulsion. Hinze scale provides an estimate for the average domain size. For a binary-fluid mixture, in comparison to a single-component fluid, kinetic energy content is suppressed. We show that this is because the presence of interfaces opens up an alternate kinetic energy transfer mechanism. Emulsion domains absorb kinetic energy up to Hinze scale, dissipate part of it and redistribute the rest to small scales.  Surprisingly, even with an alternate energy transfer mechanism, we do not find any qualitative change in the statistics of small-scale structures.

Our results have striking similarity with those of polymeric turbulence \citep{per06,per10,val14}. There also the presence of polymers modifies the energy transfer in the inertial range.  The similarity could be attributed to a close correspondence of the NSCH equations, and the equations for uniaxial polymers  \citep{bal04}. However, a crucial difference is that the  typical size of the polymer is in the dissipation range and they are homogeneously distributed throughout the flow. Whereas the size of  binary-fluid domains lie within the inertial range  \citep{aro84,per14,fan16}. 

\vspace{2cm}
{\bf Acknowledgements} \\
We thank M.M. Bandi, M. Barma, D. Mitra, and F. Toschi for their comments and suggestions.

%%\bibliographystyle{plainnat}
%%\bibliography{turb_ref}
\end{document}

%% file: symdef.inc
\def \Rey  {\mbox{Re}}
\def \We  {\mbox{We}}

\newcommand{\beq}{\begin{equation}}
\newcommand{\eeq}{\end{equation}}
\DeclareMathOperator{\Tr}{Tr}

\newcommand{\REM}[1]{{}}

\newcommand{\Eq}[1]{Eq.~(\ref{#1})}
\newcommand{\Fig}[1]{Fig.~(\ref{#1})}
\newcommand{\subfig}[2]{Fig.~(\ref{#1}#2)}